\documentclass[3p]{elsarticle}

\usepackage{setspace} 
\usepackage{textcomp}	 

\usepackage{amsmath} 	  

\usepackage{booktabs}    

\usepackage{graphicx}    

\usepackage{subfig}	 

\usepackage{doi} 

\journal{Journal of The Electrochemical Society}

\begin{document} 

\begin{frontmatter}

\title{Using a dual plasma process to produce cobalt--polypyrrole catalysts for the oxygen reduction reaction in fuel cells -- part I: characterisation of the catalytic activity and surface structure}

\author[INP]{Christian Walter\corref{cor1}}
\ead{walter@inp-greifswald.de}
\address{phone: +49 3834 554416
fax : +49 3834 554301}

\author[ESRF]{Kurt Kummer}

\author[TUD]{Denis Vyalikh}

\author[INP]{Volker Br\"user}

\author[INP]{Antje Quade}

\author[INP]{Klaus-Dieter Weltmann}

\cortext[cor1]{Corresponding author}

\address[INP]{
Leibniz Institute for Plasma Science and Technology,
INP Greifswald e.V.,
Felix-Hausdorff-Str. 2,
17489 Greifswald,
Germany
}

\address[ESRF]
{European Synchrotron Radiation Facility, 
6 Rue Jules Horowitz, 
B.P. 220, 
F-38043 Grenoble Cedex, 
France}

\address[TUD]{
Institut f\"ur Festk\"orperphysik,
TU Dresden, 
Zellescher Weg 16,
01062 Dresden,
Germany
}

\begin{abstract}

A new dual plasma coating process to produce platinum-free catalysts for the oxygen reduction reaction in a fuel cell is introduced. The catalysts thus produced were analysed with various methods. Electrochemical characterisation was carried out by cyclic voltammetry, rotating ring- and rotating ring-disk electrode. The surface porosity of the different catalysts thus obtained was characterised with the nitrogen gas adsorption technique and scanning electron microscopy was used to determine the growth mechanisms of the films. It is shown that catalytically active compounds can be produced with this dual plasma process. Furthermore, the catalytic activity can be varied significantly by changing the plasma process parameters. The amount of H$_2$O$_2$ produced was calculated and shows that a 2 electron mechanism is predominant. The plasma coating mechanism does not significantly change the surface BET area and pore size distribution of the carbon support used. Furthermore, scanning electron microscopy pictures of the produced films are presented and show the preference of columnar growth mechanisms. By using different carbons as the support it is shown that there is a strong dependence of the catalytic activity that is probably related to the chemical properties of the carbon.

\end{abstract}


\end{frontmatter}

%

\section{Introduction}

Polymer electrolyte membrane fuel cells (PEMFCs) were recognised as a potential future power source for zero emission vehicles \cite{Bashyam2006}. Unfortunately, up to now platinum has been the only efficient catalyst for the oxygen reduction reaction (ORR) in PEMFCs. For reasons of availability and cost efficiency there is a great desire to replace platinum with inexpensive and abundant catalysts (non-noble metal catalysts, NNMCs). \cite{Sealy2008}

Metal (Co/Fe):N:C composites have emerged as the most promising alternatives here. \cite{Jaouen2009,Zhang2008b} 

There are two main routes to producing such NNMCs\cite{Zhang2008b}. The first one is to start from a macrocycle containing (Co/Fe)--N$_4$ moites (e.g. porphyrins). This macrocycle is then pyrolysed at a certain temperature to increase its activity and stability \cite{Zagal2006,Bogdanoff2004,Charreteur2009}.

It was already shown that -- in some cases -- a plasma treatment instead of pyrolysis can enhance the catalytic activity even further\cite{Herrmann2010}.

However, since such macrocycles are rather expensive, another way of creating NNMCs has been developed in recent years. In this approach the metal (Co/Fe), carbon and nitrogen are introduced separately from different precursors. By pyrolysis of these mixed compounds, catalytically active bonds are formed and so cheap NNMCs can be created. But still even the best catalysts obtained by this method cannot yet compete with platinum when it comes to activity and stability yet. \cite{Jaouen2009,Barazzouk2009} 

A reason for this could be the fact that -- due to the high temperatures used in those processes -- there is an upper limit to the metal content that can be incorporated into the carbon/nitrogen structure to obtain catalytically active compounds. \cite{Lef`evre2000,Jaouen2007} Increasing the metal content further than that only leads to the formation of non-active metal particles \cite{Goubert-Renaudin2011,Lalande1997,Faubert1999,Zagal2006} due to the high mobility of the metal atoms in the heating process. \cite{Yang2007,Yang2007a}  Another reason might be the fact, that the carbon support of the catalyst changes its surface when heated to such high temperatures. Since the surface structure and composition has a strong influence on the catalytic activity \cite{Jaouen2009} a change of this structure could also be a limiting factor for the catalytic activity.

It is known that by using low temperature plasma processes, it is possible to obtain similar chemical reactions to those in the high temperature regime with the advantage of having a low sample temperature at the same time. This is, because in low temperature plasma processes only the electrons are heated up to 1\,eV--5\,eV, whereas the heavy particles (neutrals, radicals and ions) stay close to room temperature \cite{Lieberman2005} These highly energetic electrons are then able to dissociate, ionise and excite the gas molecules so that for example the polymerisation of various monomers is possible (plasma enhanced chemical vapour deposition -- PECVD). 

Physical vapour deposition (PVD) is another vacuum depositing method to deposit thin films -- mostly metals -- by the condensation of a vapourised target. PVD methods include e.g. electron beam deposition, pulsed laser deposition or sputter deposition. Sputter depositon is also a low temperature plasma process, where a voltage is applied to the target electrode, so that ions generated in the plasma are accelerated towards it. When the ions are hitting the surface, collision cascades are created which can lead to atoms being sputtered from the target. With this technique, thin films can be deposited on a substrate at low temperature. 

Because of the advantages of low temperature plasma processes, this method is currently being widely investigated for the preparation of different catalysts. \cite{Liu2002b} For that reason it is promising to use such a process for the production of active NNMCs while suppressing the formation of non active species by keeping the substrate temperature -- and thus the molecule mobility -- low. Furthermore -- by using the low sample temperatures in a plasma process -- it might be possible to keep the microstructure of the substrate carbon powder.

To produce active catalysts, powderous substrates have to be coated. The modification of powders in a plasma environment to tailor particles with desired surface properties is under intensive research. E.g. microparticles can be charged in a plasma (so called dusty plasmas) leading to a vast field and new applications of plasma-processed particles. \cite{Kersten2001,Kersten2003a,Kersten2003b}

Cobalt and iron based catalysts for the oxygen reduction reaction were already prepared by a combinatorial plasma sputter process by Easton et al. \cite{Easton2006, Easton2008, Yang2007, Yang2007a} This process consists of a pure PVD-process, where the metal is sputtered at the same time as carbon in a nitrogen atmosphere to obtain (Co/Fe):N:C-compounds. However, those composites only show activity for the oxygen reduction reaction when they are heated to temperatures above 650\,\textcelsius\ after the plasma process, leading to similar problems as encountered with purely pyrolised samples. Furthermore, they observe that the nitrogen content within their samples decreases with increasing heat treatment temperature -- nitrogen is lost by the heat treatment -- leading to a loss of catalytically active sites.

In this contribution, we try to overcome this conflict by using a dual PECVD/PVD process instead of a combinatorial sputtering process to produce catalytically active Co:N:C complexes. In this process, a plasma polymerisation process is coupled with the sputtering of the metal to obtain catalytically active compounds with the advantage, that a carbon and nitrogen containing monomer can be used for the production of Co/Fe--N--C structures. So, the backbone of the NNMC already exists in the precursor and only the metal has to be incorporated to this structure in a convenient way.

It has already been shown that metal--polymer-composites with versatile properties can be produced in such a plasma process. \cite{Walter2009,Korner2010, Korner2011} To produce NNMCs, we used the dual PECVD/PVD process with pyrrole as the nitrogen precursor and cobalt as the sputtered metal since it has already been shown that catalytically active compounds can be produced with cobalt and pyrrole chemically. \cite{Bashyam2006,Mill'an2009,Yuasa2005}

In this contribution we show that catalytically active compounds can also be produced with a dual plasma process only -- without any heating of the substrate. Furthermore, by tuning the plasma process parameters, the catalytic activity can be varied significantly. In the first part of the paper, the films thus obtained are analysed electrochemically by cyclic voltammetry (CV), rotating-disk electrode (RDE) and rotating-ring-disk electrode (RRDE). The topological structure is analysed by means of scanning electron microscopy (SEM) and the surface structure and pore size distribution is investigated with the gas adsorption technique. Furthermore, different carbon supports are used as a substrate to determine their influence on the catalytic activity of the compound.

In the second part of the paper, the structure of the obtained NNMCs is investigated to get information on the chemical structure -- and thus the nature of the probable catalytically active sites -- of the compounds. This is done by several surface-sensitive methods -- X-ray photoelectron spectroscopy (XPS) and near edge X-ray absorption spectroscopy (NEXAFS) -- as well as bulk-sensitive methods -- energy-dispersive X-Ray Spectroscopy (EDX), X-ray diffraction (XRD) and extended X-ray absorption fine structure (EXAFS).

\section{Materials and Methods}

\begin{figure*}[htb]
	\centering
	\includegraphics[width=\linewidth]{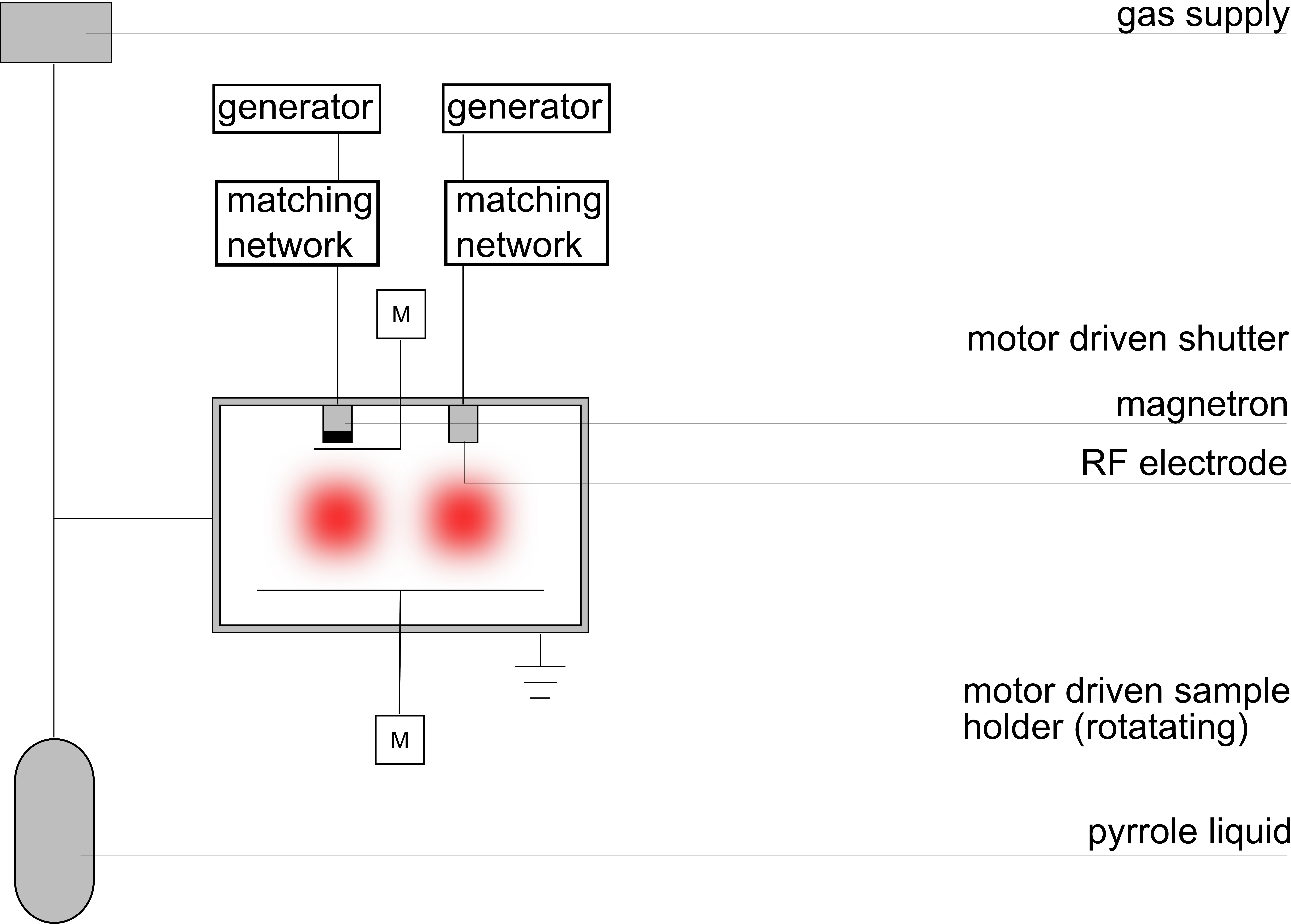} 
	\caption{Schematic illustration of the recipient used for the production of cobalt--polypyrrole films\protect\footnotemark}
	\label{fig:recipient}
\end{figure*}

As schematically shown in figure \ref{fig:recipient}, plasma polymerisation of pyrrole (C$_4$H$_5$N) was carried out within a rectangular shaped recipient with a ground area of 400\,mm*400\,mm and a height of 200\,mm. 50\,sccm argon controlled by a flow-meter and evaporated pyrrole (Sigma-Aldrich; reagent grade) controlled by a needle valve were introduced into the chamber. The process was then carried out at a constant pressure of 7\,Pa (controlled by a butterfly valve).

For the production of cobalt films, an RF magnetron with cobalt target (target height 1\,mm) was installed in the chamber. Pure argon (50\,sccm) was used as the sputter gas at a pressure of 7\,Pa. The power of the magnetron was varied between 25\,W and 400\,W leading to an increase of self-BIAS-voltage from 100\,V to 700\,V.

For the production of cobalt--polypyrrole films, both plasma processes were coupled. The RF-electrode was used for the polymerisation of the pyrrole monomer (PECVD-process) and the magnetron was used to incorporate cobalt into those films (PVD-process). To ensure a uniform coating of the substrate with both cobalt and polypyrrole, the sample holder was rotated at a distance of 50\,mm to the electrodes. The substrate was kept on floating potential -- aprox. 10\,V, slightly depending on the plasma-power; measured with \emph{Smart Probe}\footnote{Straatum Processware Ltd.} -- during the film growth. To avoid poisoning of the target, both processes were carried out stepwise (alternatingly) and pyrrole was not present within the recipient while the magnetron was turned on. Furthermore, the shutter in front of the magnetron was closed  during the PECVD-process so that the magnetron is not coated with polypyrrole. To effectively produce a cobalt--polypyrrole compound and prevent a sequential structure of cobalt and polypyrrole films, the time used for one step was short (9 sec) while the steps were carried out up to 50 times to increase the film thickness.
 
For all the samples presented in this paper, the power of the RF-electrode was held constant at 40\,W while the magnetron power was varied.

Measurements of the catalytic activity were performed by cyclic voltammetry (CV) in a solution of sulphuric acid of pH\,1, saturated with pure oxygen. To ascertain that the values obtained there actually arise from the oxygen reduction reaction and that other electrochemical reactions can be excluded, a test measurement in inert nitrogen has always been carried out before the oxygen measurement. The catalysts are produced by depositing the cobalt--polypyrrole films onto carbon powder -- when not stated otherwise, Cabot Vulcan XC72R is used as support. An ink is prepared by mixing 2\,mg of the powder thus produced, 20\,$\mu$l of 5\,wt.-\% Nafion solution (Sigma Aldrich) and 70$\,\mu$l of ethanol (Sigma Aldrich) for 15\,min in an ultrasonic bath. Seven microliters of ink are then pipetted onto a glassy carbon disk (0.196\,cm$^2$), resulting in a catalyst loading (carbon powder plus cobalt--polypyrrole film) of 0.8\,mg\,cm$^{-2}$. Cyclic voltammograms with a sweeping rate of 10\,mV\,s$^{-1}$ are recorded in the potential range of -0.2\,V to 0.8\,V versus a saturated Ag/AgCl electrode. In the downward potential scan, a reduction current peak occurs, owing to O$_2$ reduction kinetics and O$_2$ depletion. This peak position (V$_\mathrm{PR}$) has been shown to be correlated both theoretically and experimentally with the exchange current density in a fuel cell \cite{Jaouen2003,Lef`evre2000,M'edard2006} and is thus a good measure for the catalytic activity. All voltages reported in the results section of this paper are given with respect to RHE. The potential used for the conversion of saturated Ag/AgCl to RHE is 0.22\,V. \cite{Bates1978}

Additionally, Tafel-plots were obtained from rotating ring electrode (RDE) measurements carried out at a rotation speed of 3500\,rpm. The kinetic current density $I_\mathrm{kin}$ was calculated by $I_\mathrm{kin}=I_\mathrm{lim}*I_\mathrm{D}/(I_\mathrm{lim}-I_\mathrm{D})$, where $I_\mathrm{D}$ is the measured disk current density and $I_\mathrm{lim}$ the limiting current density (taken at 0\,V vs RHE).

Rotating ring-disk electrode (RRDE) measurements were performed with the same setup as described in the CV section. The disk diameter is 5\,mm, and the ring inner and outer diameters are 6.5\,mm and 7.5\,mm respectively. Before each measurement the platinum ring was cleaned by cycling it 10 times from -0.3\,V--1.3\,V vs. Ag/AgCl to remove contaminations that could lead to a decrease of the H$_2$O$_2$ oxidation activity of platinum. \cite{Marcotte2004} The measurements were carried out at a rotation speed of 200\,rpm. This falls within the range for RRDE measurements where laminar flux can be ensured. \cite{Bard2000} The disk was then cycled from 0.8\,V to -0.2\,V vs. Ag/AgCl while the platinum ring was held constant at 1.2\,V vs. Ag/AgCl to be able to completely oxidise the produced H$_2$O$_2$. The scan in O$_2$ was then corrected by the non-faradayic current that was measured with the same procedure in N$_2$ saturated electrolyte before each measurement. After this, the number of electrons transferred was calculated with the formula

\begin{equation}
n=\frac{4I_\text{D}}{I_\text{D}+(I_\text{R}/N)}
\end{equation}

where $I_\text{D}$ and $I_\text{R}$ are the ring and disk current. \cite{Zhang2008b} $N$ is the collection efficiency of the ring. It was determined experimentally with the Fe(II)/Fe(III) redox couple -- 2\,mmol/l K$_3$(Fe(CN)$_6$)solved in 0.1\,mol/l KNO$_3$ -- to be 0.24 (fits well to the theoretically calculated value of 0.26).  After this, the percentage of H$_2$O$_2$ produced was calculated by

\begin{equation}
\% \text{H}_2\text{O}_2=100\cdot\frac{4-n}{2}.
\end{equation}

\cite{Zhang2008b}

Scanning electron microscope (SEM) pictures were taken using a field emission scanning electron microscope JEOL JSM-7500 F with a resolution of 1\,nm at 15\,kV acceleration voltage. The films were therefore deposited onto a silicon (100) surface. Before each measurement, the surface was scratched with a scalpel to be able to get a cross-sectional picture of the film.

The measurements of the BET surface and the micro- and mesopores of the pure and NNMC-coated carbon powders were carried out using a Quantachrome Nova 2200 device. Nitrogen was used as the adsorption gas and the samples were cooled to liquid nitrogen temperatures. The pore size distribution was determined using quenched solid state density functional theory (QSDFT) with a slit-pore model. This model improves the accuracy of disordered carbon materials because it explicitly takes into account the effect of surface roughness and heterogeneity explicitly. \cite{Instruments} For the BET and pore-size distribution measurements the films were deposited on different carbon powders. The film thickness was adjusted in such a way that the lowest thickness possible -- without loss of catalytic activity -- was used.

\section{Results and Discussion}

\subsection{Cyclic Voltammetry}

\begin{figure*}[htb]
	\centering
	\includegraphics[width=\linewidth]{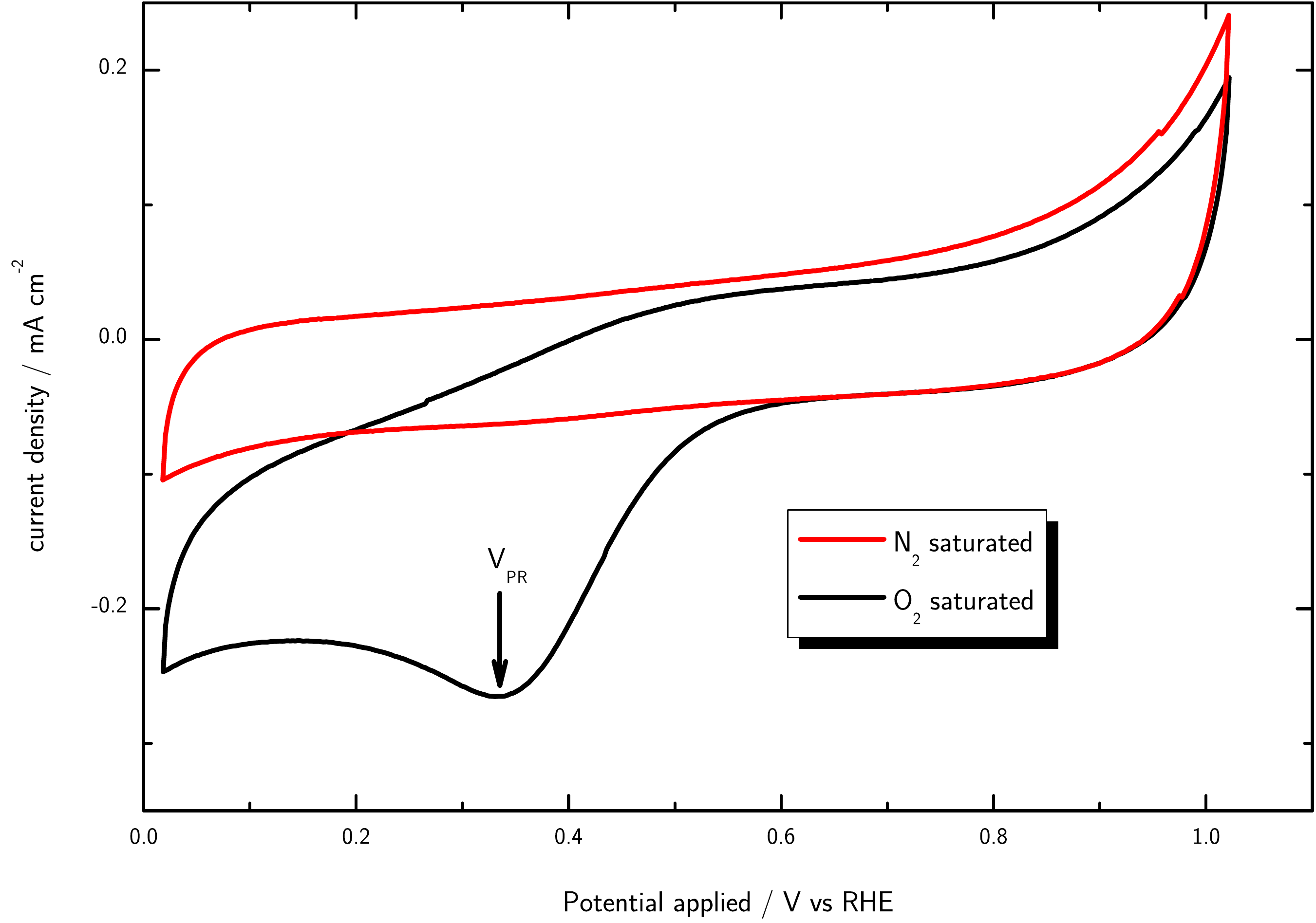} 
	\caption{CV measurement of cobalt--polypyrrole produced with 400\,W magnetron power in H$_2$SO$_4$ at pH\,1; the position of the oxygen reduction peak is shown and marked as V$_\mathrm{PR}$}
	\label{fig:CV}
\end{figure*}

Exemplarily, the CV-measurement of a sample produced with a magnetron power of 400\,W is shown in figure \ref{fig:CV}. As can easily be seen, the reference measurement in inert nitrogen does not show any peaks. This is a proof that no redox reaction is taking place within the film itself. By measuring the same film in oxygen saturated sulphuric acid a peak evolves in the reduction branch. This peak -- also marked in figure \ref{fig:CV} -- is assigned to the oxygen reduction reaction taking place at the cobalt--polypyrrole layer. This position is directly correlated with the exchange current density in a fuel cell \cite{Jaouen2003,Lef`evre2000,M'edard2006}. Hence, this peak position is used to characterise the catalytic activity of the plasma-produced samples within this paper.

In table \ref{tab:CV} the peak positions of cobalt--polypyrrole films produced with different magnetron powers are given. It is shown that they increase with raising magnetron power up to a maximum value of 0.34\,V. This value is comparable to the values obtained with chemically produced cobalt--polypyrrole films (0.3\,V). \cite{Mill'an2009}

By comparing the catalytic activity with other -- chemically produced with other precursors than cobalt and pyrrole -- compounds it is clear that the plasma produced samples are still not as good as the best of those. The group of Dodelet reached values up to about 0.6\,V V$_\mathrm{PR}$ with pyrolised cobalt--acetate. \cite{Marcotte2004} With sol-gel-derived Co:N:C-based catalysts, values up to 0.58\,V have been achieved. \cite{Sirk2008}  

With pretreated carbon supports impregnated with 0.2\,wt.-\% iron-acetate and heated up to 950\,\textcelsius\ values up to as high as 0.7\,V have been reached. \cite{Barazzouk2009}

The increase in catalytic activity with raising magnetron power is probably due to the higher amount of Co:N:C structures that form in these samples. As will be shown in part II of this paper, a higher magnetron power leads to an increase of cobalt within the samples because of a higher cobalt sputter yield with raising magnetron-power. The energy of the of the cobalt atoms arriving at the substrate is also increased. This leads to the formation of more and more Co:N:C structures. These are known to be catalytically active for the oxygen reduction reaction. 

The magnetron power can -- in principle -- be raised above the value of 400\,W but due to the specifications of the magnetron we used, higher powers could not yet be achieved. So by using higher magnetron powers the catalytic activity can probably be enhanced further.

\subsection{Rotating disk electrode}

\begin{figure*}[htb]
	\centering
	\includegraphics[width=\linewidth]{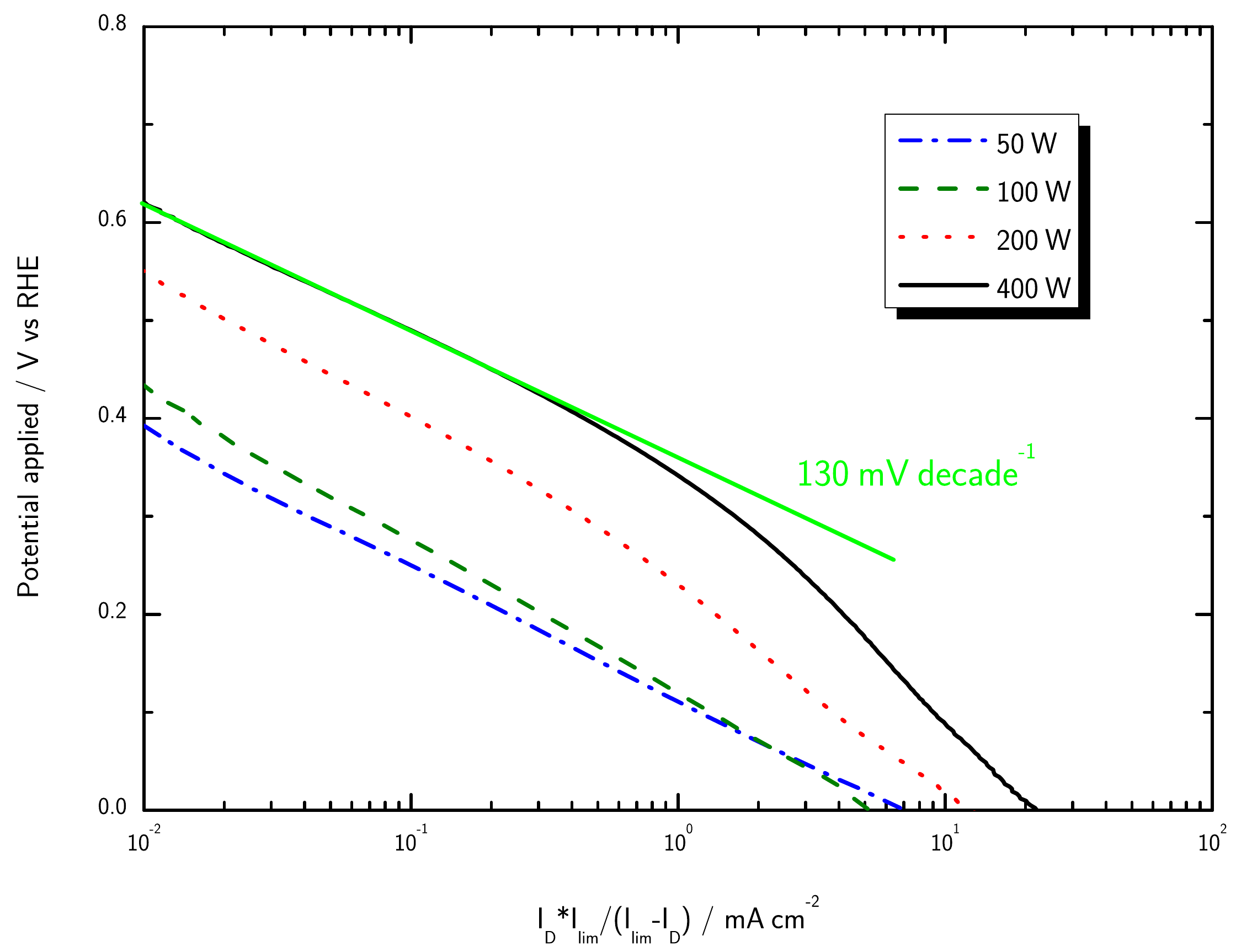} 
	\caption{Tafel-plots of cobalt--polypyrrole composites produced with different magnetron powers}
	\label{fig:tafelplot}
\end{figure*}

The Tafel-plots for the composites produced with varying magnetron power are given in figure \ref{fig:tafelplot}. At a voltage of 0.4\,V the current density at the cobalt--polypyrrole-composite, produced with a magnetron power of 50\,W, is 0.01\,mA\,cm$^{-2}$. It increases up to 0.45\,mA\,cm$^{-2}$ at the compound produced with a magnetron power of 400\,W. This confirms the increasing catalytic activity of the compounds with raising magnetron power that was already observed with cyclic voltammetry. The comparison of this results to the ones obtained for the best NNMCs can not be done by studying the current densities at a voltage of 0.4\,V vs RHE since most of the results reported are not given at this voltage. Therefore, the differences are shown by comparing the voltages at which a current density of 0.45\,mA\,cm$^{-2}$ can be found.  

In cobalt porphyrins, current values of 0.45\,mA\,cm$^{-2}$ were observed at a electrode potentials of 0.7\,V vs RHE. \cite{Olson2009,Bouwkamp-Wijnoltz1999} By adding sulfur to the pyrolysis of the compounds, even values up to 0.8\v vs RHE were observed. \cite{Herrmann2009a} With iron containing compounds, 0.85\,V were reached. \cite{Jaouen2009a}

The Tafel-slope of the compounds at low overpotentials is 130\,mV\,decade$^{-1}$ and does not change with varying magnetron power. The similar Tafel-slopes indicate, that the reaction mechanism is the same for all catalysts. For the best NNMCs thus obtained, Tafel-slopes in the range of 54\,mV\,decade$^{-1}$--65\,mV\,decade$^{-1}$ -- similar to the 60\,mV\,decade$^{-1}$ that are reported for platinum catalysts -- were observed. There, a direct 4 electron process is proposed. \cite{Jaouen2009} At catalysts with a lower current density, Tafel-slopes up to 200\,mV\,decade$^{-1}$ are found. \cite{Jaouen2006a} The higher Tafel-slopes are probably related to a different reaction mechanism of the ORR. \cite{Jaouen2003} So, the reaction mechanism of the compounds produced with the dual plasma process is probably not the favoured 4 electron reaction, but similar to the catalysts with lower activity. This also correlates with the fact, that the catalytic activity of the plasma produced compounds is not as good as the best values reported.

\subsection{Rotating ring-disk electrode}

\begin{figure*}[htb]
	\centering
	\includegraphics[width=\linewidth]{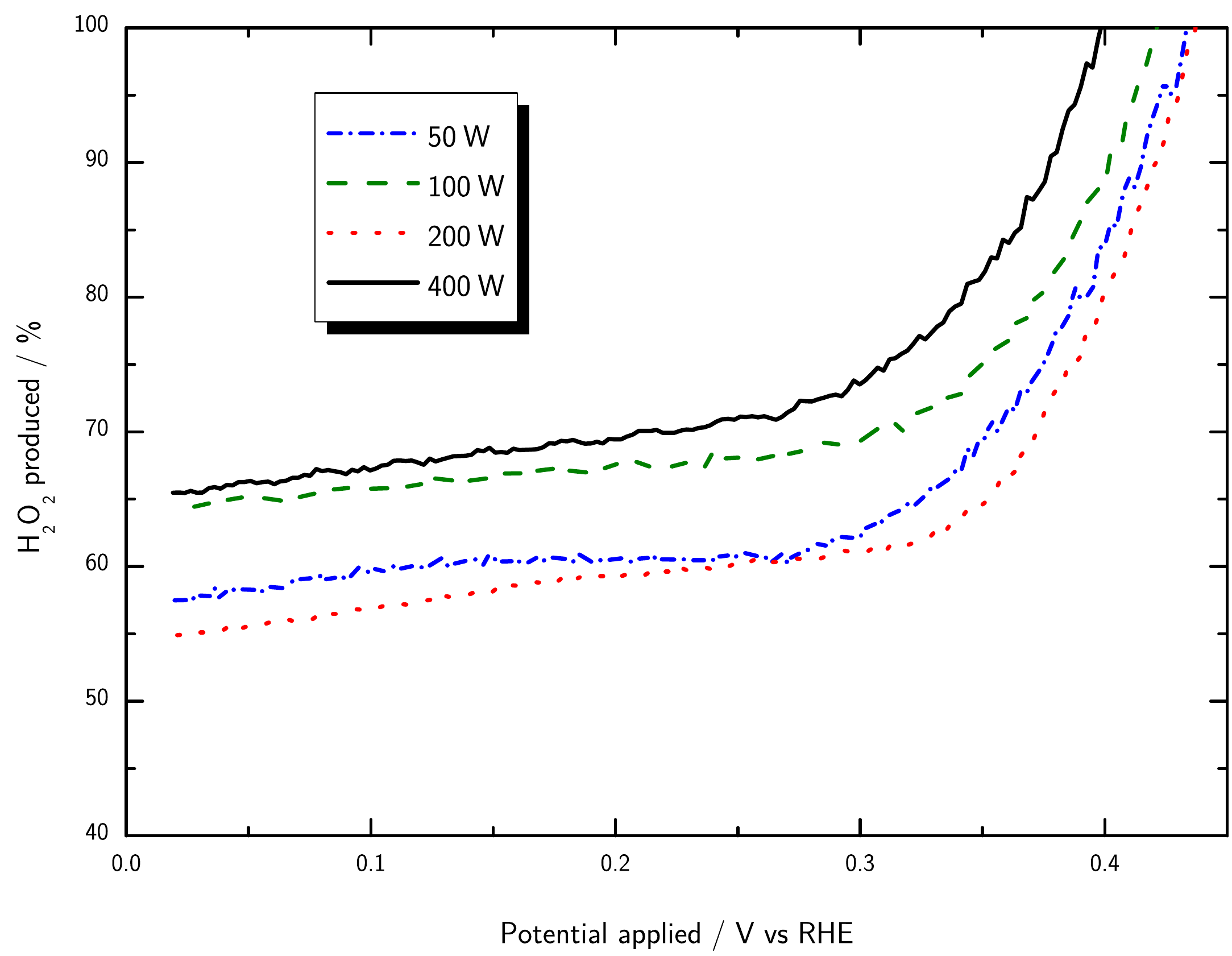} 
	\caption{H$_2$O$_2$ generated at cobalt--polypyrrole produced with different magnetron powers}
	\label{fig:RRDE}
\end{figure*}

The percentage of H$_2$O$_2$ produced at the electrodes for different magnetron power values, calculated at a rotation speed of 200\,rpm, is shown in figure \ref{fig:RRDE}. No clear trend can be observed with varying magnetron powers but the qualitative behaviour of all samples is the same. This can be correlated to the constant Tafel-slope at all samples and confirms, that the oxygen reduction mechanism is the same at all compounds. The percentage of H$_2$O$_2$ produced decreases down to about 60--70\,\% at 0.2\,V vs. RHE. This decrease is characteristic for thin film electrodes where a 2 electron reaction + H$_2$O$_2$ electroreduction mechanism is present. \cite{Jaouen2009b} So, not the favoured 4 electron reaction -- O$_2$ + 2 H$_2$ $\rightarrow$ 2 H$_2$O -- but rather the 2 electron reactions -- O$_2$ + H$_2$ $\rightarrow$ H$_2$O$_2$ and H$_2$O$_2$ + H$_2$ $\rightarrow$ 2 H$_2$O -- take place at the catalysts produced with the dual PECVD-PVD process. \cite{Jaouen2009a,Jaouen2009b} 

This is also in agreement with the conclusions from the Tafel-plots. A direct 4 electron reaction leading to a Tafel-slope of approx. 60\,mV\,decade$^{-1}$ -- as in the most catalytically active NNMCs -- is not observed. The high Tafel-slope of 135\,mV\,decade$^{-1}$ is therefore probably related to the different ORR mechanism of the 2 electron reactions. To enhance those catalysts further, it is therefore necessary to improve the catalytically active structure so that a direct 4 electron process is present.

\subsection{Pore size distribution}

\begin{figure*}[htb]
	\centering
	\includegraphics[width=\linewidth]{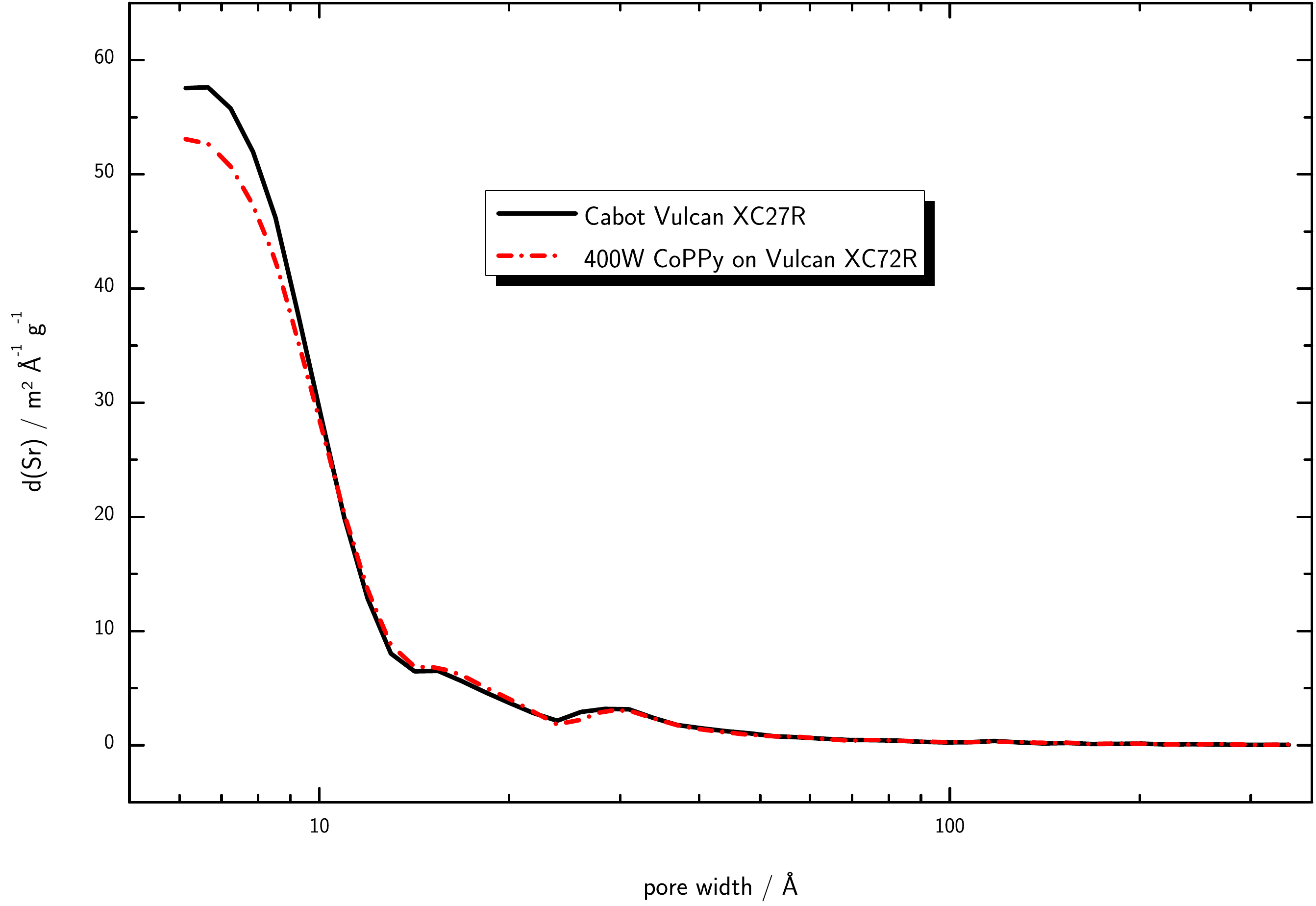} 
	\caption{Differential surface area distribution obtained by analysing the N$_2$ isotherms with the QSDFT model}
	\label{fig:poresize}
\end{figure*}

The pore size distribution of pure and cobalt--polypyrrole coated Cabot Vulcan XC72R is given in figure \ref{fig:poresize}. As can be seen, the pore size distribution does not vary strongly. The only change that can be seen is that the amount of pores smaller than 10\,\AA\ decreases slightly on the coated samples. This can easily be explained by the coating mechanism used in this paper. Since the carbon powder is coated with plasma processes it can be expected that the smallest pores will be filled up first -- leading to a decrease of those pores on the carbon powder. When the process is carried out for longer time-scales, it can be anticipated that also larger pores will be filled up with catalyst which might lead to a decrease of mesopores in the film. But since no further enhancement in catalytic activity could be observed by longer treatment of the carbon support, such a long coating time period was not used.

However, since no drastic change can be observed at Vulcan XC72R, it can be concluded that the dual PECVD/PVD process presented here can avoid the changes that are usually accompanied by the heating process of the catalysts that is normally used for the production of NNMCs.

\subsection{Using various carbon substrates}

It has been shown with chemically produced samples that one important factor for the catalytic activity of the NNMCs is the ability of the carbon support to bind nitrogen -- represented by the nitrogen content at the surface of the carbon support. \cite{M'edard2006,Jaouen2003,Marcotte2004} Furthermore, it has also been reported that the disordered phase of the carbon \cite{Jaouen2006a} and the microporous area of the catalyst \cite{Jaouen2009,Yang2009} influence the catalytic activity.

To check if such effects can also be observed at plasma-produced samples, different carbon supports were used as substrate and the surface structure as well as the catalytic activity was measured. The summary of the micro- and mesoporous as well as the BET surface and the catalytic activity of catalysts deposited onto different carbon powders is presented in table \ref{tab:BET}. As can be seen, the catalytic activity differs strongly depending on the carbon substrate used. The best value of 0.41 is obtained with Norit SX Ultra. As was already shown for the pore size distribution on Vulcan XC-72R explicitely, no strong change of the carbon substrate pore areas can be seen here either. Only the microporeous area decreases slightly at all of the samples, the BET-area remanains nearly constant and the mesoporeous area increases slightly. As has already been explained, this could be related to the plasma coating process. Micropores are filled up first, leading to a decrease of micropores and an increase of mesopores at the coated samples.

No correlation can be found between either BET, micro- or mesoporous area and the catalytic activity thus obtained. The reason for the difference in catalytic activity is therefore probably related to different chemical compositions of the carbon -- leading to different depositing mechanisms in the plasma process. The nitrogen-binding ability of the carbon surface support probably plays the most important role. So to further optimise the catalysts the use of other carbon supports will probably be helpful. 

Furthermore, it has already been observed that a pretreatment -- pyrolysis of the pure carbon support in NH$_3$/H$_2$/Ar -- can increase the catalytic activity of chemically produced samples drastically. \cite{M'edard2006,Jaouen2003,Marcotte2004} The catalytic activity of the plasma-produced catalysts can probably also be enhanced by a pretreatment of the carbon support. 

Last but not least, it has also been shown that a pretreatment in an N$_2$ plasma leads to an enrichment of N-groups on the carbon surface, leading to a reduction of the H$_2$O$_2$ production during the catalysis. \cite{Herranz2007} Since the catalysts presented in this paper show high H$_2$O$_2$ production rates such a plasma pretreatment migth help in gaining catalysts that show a 4 electron reduction pathway. This could lead to a higher overall catalytic activity of the samples thus produced. Since the catalyst is prepared in a dual plasma process anyway, the pretreatment can easily be carried out in the same recipient before the catalyst coating process -- avoiding the need to put the sample to air atmosphere -- and thus probably oxidising the sample again. 

\subsection{Scanning electron microscope}

\begin{figure*}[htb]
  \centering
    \subfloat[400W low magnification]{
    \label{fig:SEM_400W_lowmag}
    \includegraphics[height=0.295\textheight]{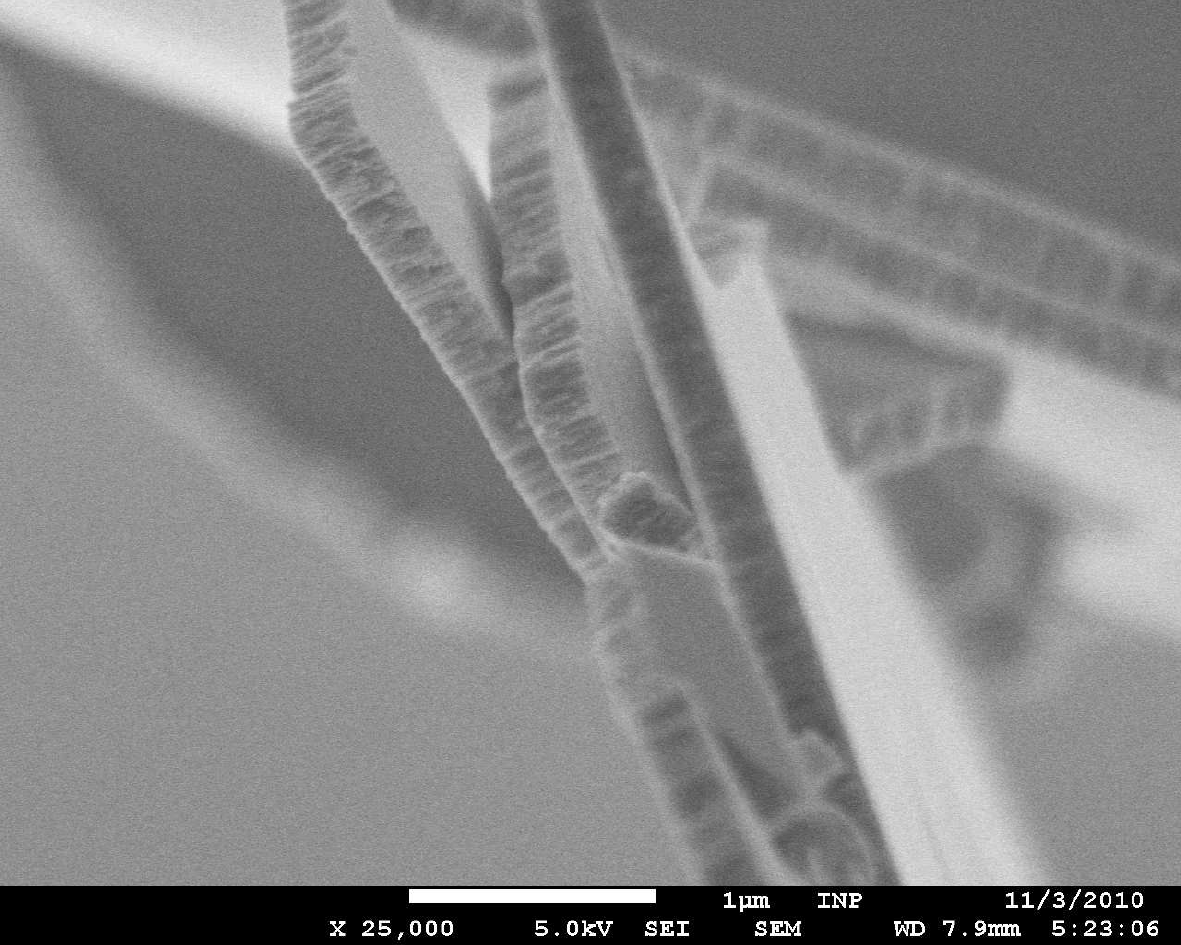} 
  }
  \subfloat[400W high magnification]{
    \label{fig:SEM_400W_highmag}
    \includegraphics[height=0.295\textheight]{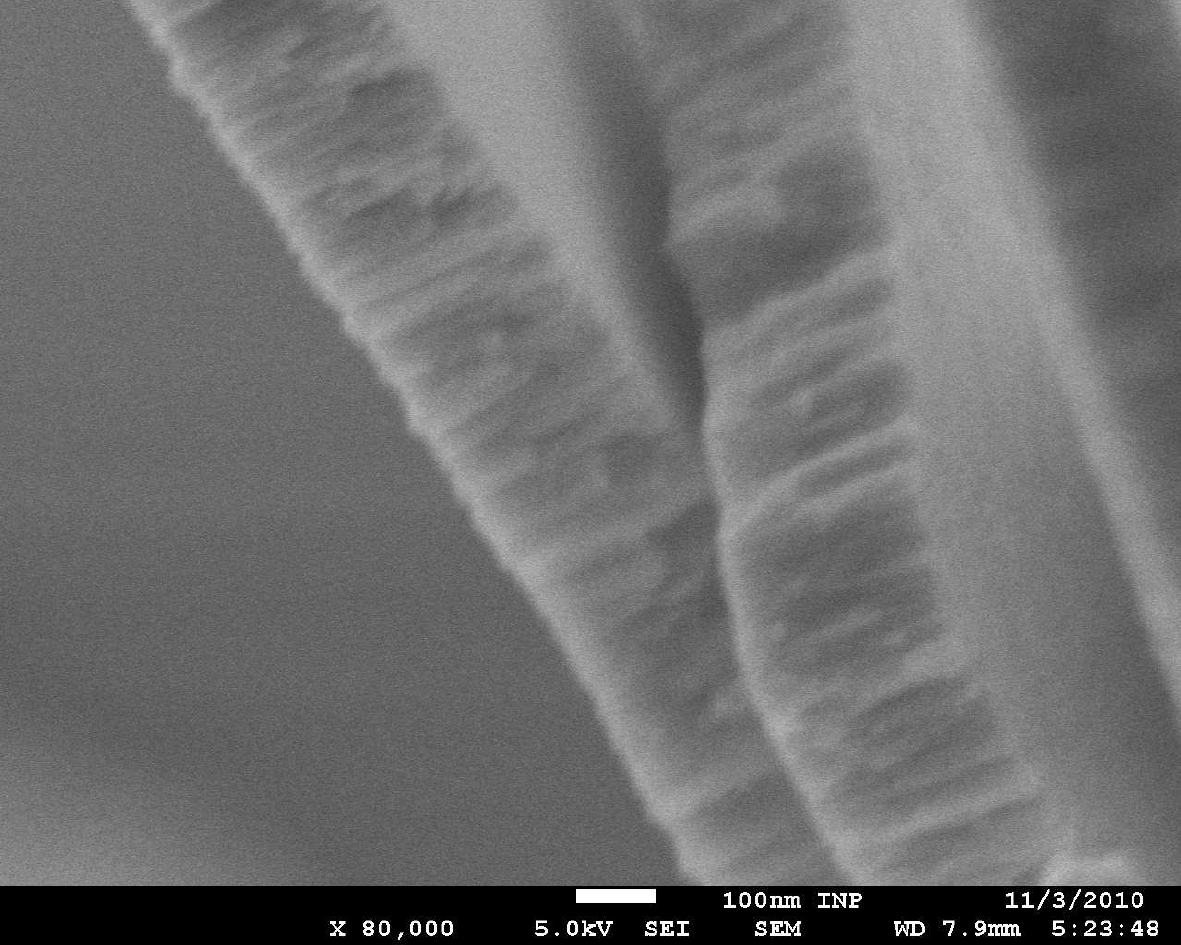} 
  }
  \\
  \subfloat[100W low magnification]{
    \label{fig:SEM_100W_lowmag}
    \includegraphics[height=0.295\textheight]{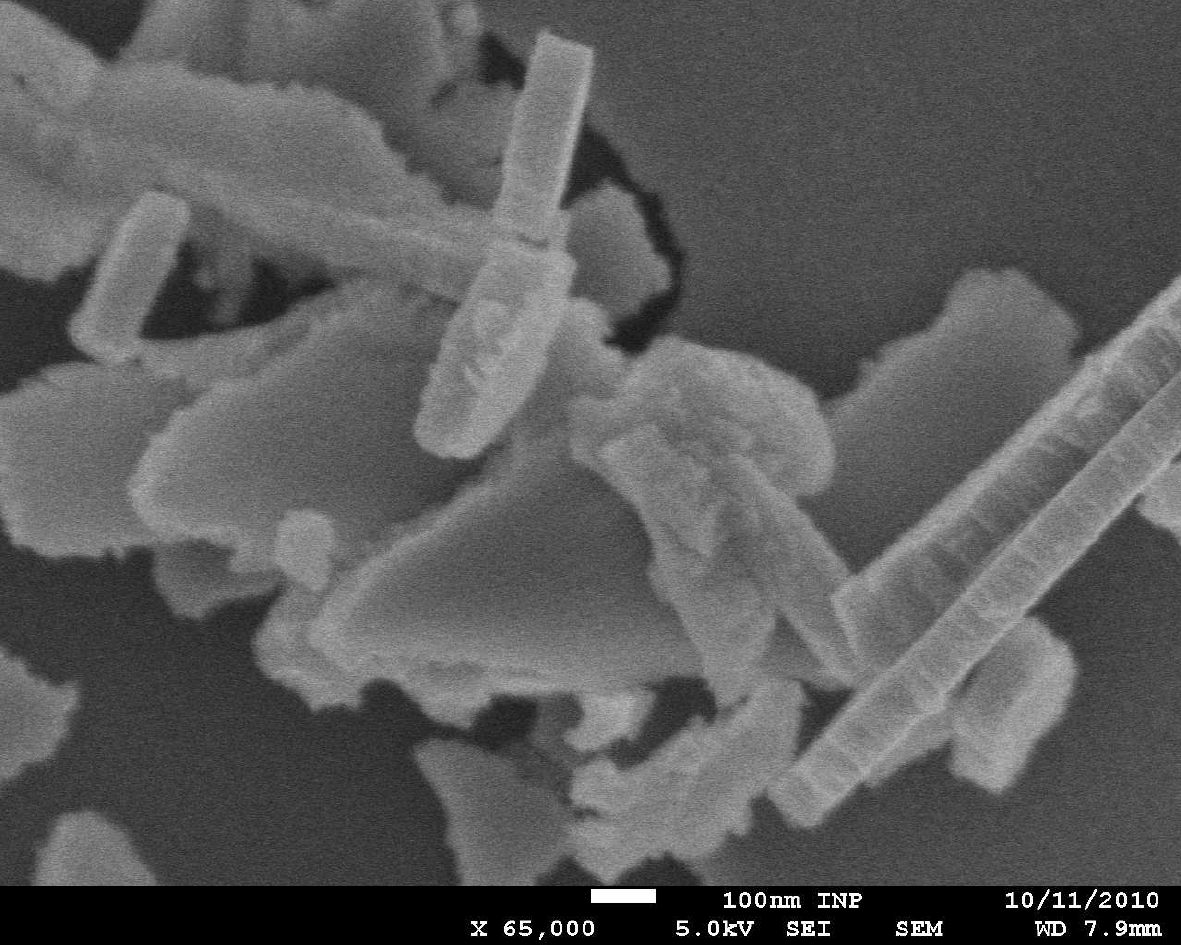} 
  }
  \subfloat[100W high magnification]{
    \label{fig:SEM_100W_highmag}
    \includegraphics[height=0.295\textheight]{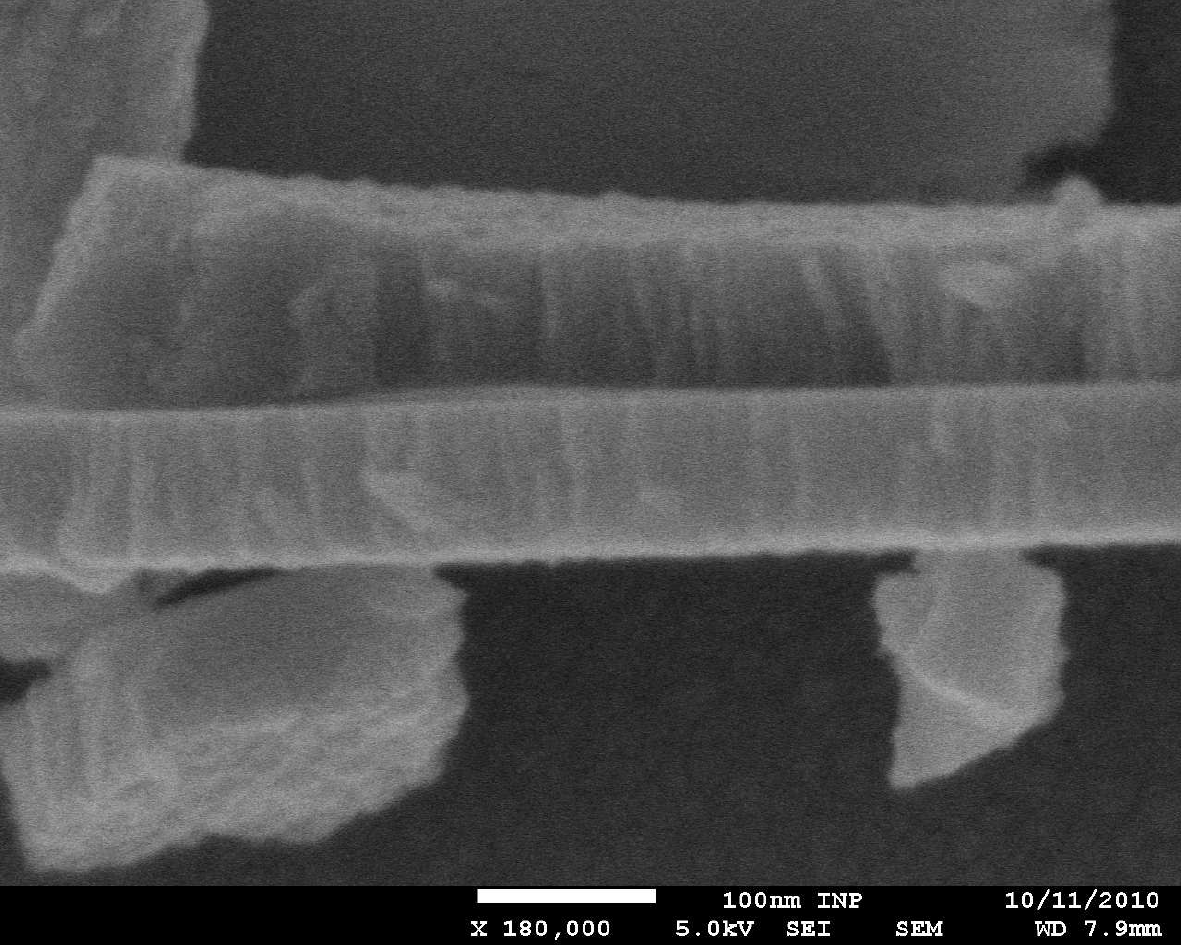} 
  }
  \caption{SEM pictures of cobalt-polypyrrole samples; cross-sectional view}
  \label{fig:SEM}
\end{figure*}

SEM pictures taken for the samples produced with a magnetron power of 100\,W and 400\,W can be seen in figure \ref{fig:SEM}. High as well as low magnification images of a cross sectional view of the films are shown. The pictures were achieved by scratching -- and therefore breaking up -- the films to be able to get a side view onto the growth structures. In all films, columnar growth is favoured. This is typical for low temperature plasma processes where the coating flux arrives on a smooth surface in a direction that is largely normal to the substrate surface so that shadowing effects are minimised. \cite{Thornton1977} Of course, the growth mechanism on the carbon powder itself is different since it has a much higher roughness and different chemical composition.

The thicknesses of the films are 90\,nm and 270\,nm for the films produced with a magnetron power of 100\,W and 400\,W respectively. This is because higher magnetron powers result in an increasing growth rate of the film due to raising sputtering yield.

\section{Conclusion}

It was shown that catalytically active cobalt--polypyrrole samples could be produced by a dual plasma process only -- without additional heating of the substrate. Furthermore, the catalytic activity varied significantly when changing the power of the magnetron plasma source. Only two precursors -- pyrrole and cobalt -- were used for the production of the films, avoiding complicated multi-step processes. The activity is in the same range as that of chemically produced cobalt--polypyrrole samples.

Mainly the 2 electron reactions -- O$_2$ + H$_2$ $\rightarrow$ H$_2$O$_2$ and H$_2$O$_2$ + H$_2$ $\rightarrow$ 2 H$_2$O -- occur at the catalyst leading to a H$_2$O$_2$ production of 60\,\%--70\,\% at 0.2\,V vs. RHE.

Different carbon substrates were used for the production of the catalyst and a drastic change of catalytic activity with varying supports was observed. No correlation between catalytic activity and pore size distribution or surface area could be observed. So the difference of the catalytic activity using different carbon supports is probably related to the chemical structure of the carbon.

Furthermore, the plasma coating process does not drastically change either the surface area or the pore size distribution of the underlying carbon.

SEM pictures show that columnar growth is favoured within all films produced here.

Further increase of the catalytic activity of the plasma produced samples by tuning the plasma parameters seems possible. First of all, raising the magnetron power above 400\,W to obtain better films is -- in principle -- possible. Due to specifications of the magnetron we used, higher powers -- and so possible further enhancement of the catalytic activity -- could not be achieved. Furthermore, the gases the process is carried out with -- we used pure argon in all the samples reported in this paper -- can also be varied. By an addition of different gases to the process many plasma parameters as electron temperature and density, sputter yield, energy impact and last but not least the chemistry of the plasma process is changed. E.g. one could introduce nitrogen to the process so that more nitrogen is incorporated into the films. This might lead to a further increase of Co:N:C and graphitic nitrogen structures within the film, leading to a higher catalytic activity. Or hydrogen could be added to suppress the oxidation of the films.

A bias can also be applied to change the ion bombardment of the substrate. This is known to alter the growth mechanisms of plasma-produced films \cite{Maissel1965} so that the catalytic activity of the films might be changed. Additionally, iron can be used instead of cobalt as the sputtered metal which might lead to a higher catalytic activity -- in chemically produced compounds iron--nitrogen--carbon compounds show a higher catalytic activity than compounds containing cobalt. 

Last but not least, on the catalysts presented here the carbon support is only coated from one side -- a plasma is a direct surface coating method. Since the carbon powder is not moved during the film production, only the upper side of the powder is covered with the catalytically active film. In principle, plasma processes can also be carried out on powderous samples with e.g. a vibrating bed setup -- ensuring a uniform coating all over the powder \cite{Vahlas2006}. Since the films here were also analysed by different measurements -- needing various substrates -- such a vibration was avoided here. The results presented here are only intended to show the general feasibility of a dual plasma process to produce catalytically active films. So, a recipient where the powder can be coated uniformly from all sides --  leading to a more complicated experiment -- was not constructed for the first attempts.

But even though the powder was only coated from one side, a drastic improvement of the catalytic activity could be reached by tuning of the magnetron-power. This shows, that plasma processes are a promising technique to produce active NNMCs.

The chemical characterisation of the films and thus information on the probable catalytically active site will be given in part II of this paper.

\listoftables

\begin{table}[htb]
	\centering
	\caption{Catalytic activity, given by the oxygen reduction reaction peak at position V$_\mathrm{PR}$; the varying magnetron power is given, the RF-electrode was held constant at 40\,W}
	\begin{tabular}{rr}
	\toprule
	magnetron power  & V$_\mathrm{PR}$ vs RHE\\ 
	\cmidrule(r){1-1} \cmidrule(l){2-2}
	W  & V \\ 
	\midrule
	25 & 0.11 \\
	50  & 0.14 \\ 
	100  & 0.25 \\ 
	200  & 0.25 \\ 
	300  & 0.29 \\ 
	400 & 0.34 \\
	\bottomrule
	\end{tabular}
		\label{tab:CV} 
\end{table}

\begin{table*}[htb]
	\centering
	\caption{Surface Area and pore volume of different carbons}
	\begin{tabular}{lrrrr}
	\toprule
	sample  & BET surface & microporous area & mesoporous area  & V$_\mathrm{PR}$ \\ 
	 & & $<$ 20\,\AA\ & 20\,\AA\ $<$ x $<$ 345\,\AA\  & \\
	\cmidrule(r){2-4} \cmidrule(l){5-5}
	  & \multicolumn{3}{c}{m$^2$\,g$^{-1}$} & V \\ 
	\midrule
	Cabot Vulcan XC72R & 240 & 174 & 54 & --- \\
	400\,W CoPPy on XC72R & 230 & 168 & 52 &  0.34 \\
	Norit Cap Super & 1\,760 & 1\,081 & 233 & --- \\
	400\,W CoPPy on Cap Super & 1\,675 & 1\,050 & 262 & 0.14 \\
	Norit SX Ultra Cat & 1\,000 & 853 & 125 & --- \\
	400\,W CoPPy on SX Ultra & 1\,000 & 843 & 137 & 0.41 \\
	AkzoNobel Ketjen Black EC-300J & 870 & 581 & 200 & ---\\
	400\,W CoPPy on EC-300J & 890 & 574 & 231 & 0.16 \\
	\bottomrule
	\end{tabular} 
	\label{tab:BET}
\end{table*}

\newpage

\listoffigures

\newpage

\renewcommand{\thefootnote}{\fnsymbol{footnote}}

\footnotetext{Reprinted from \cite{Walter2011} with permission from Elsevier}

\clearpage


\bibliographystyle{plainnat}

\end{document}